\documentstyle[preprint,prl,aps]{revtex}
\begin{document}
\preprint{IUCM95-036}
\title{Thermodynamics of Quantum Hall Ferromagnets}
\author{Marcus Kasner$^{\dagger}$ and A.~H.~MacDonald}
\address{ Dept.\ of Physics,
        Indiana University, Swain Hall West 117,
Bloomington, IN 47405, USA}
\date{\today}
\maketitle

\begin{abstract}

The two-dimensional interacting electron gas at Landau level
filling factor $\nu =1$ and temperature $T=0$ is a strong ferromagnet;
all spins are completely aligned by arbitrarily weak Zeeman coupling.
We report on a theoretical study of its thermodynamic properties using
a many-body perturbation theory approach and concentrating on the
recently measured temperature dependence of the spin magnetization.
We discuss the interplay of collective and single-particle aspects
of the physics and the opportunities for progress in our understanding
of itinerant electron ferromagnetism presented by quantum Hall
ferromagnets.

\vspace{0.3cm}
\end{abstract}
\begin{flushleft} PACS numbers: 73.40.Hm, 75.10.Lp  \end{flushleft}

\newpage

The ground state of a two--dimensional electron gas (2DEG)
in the quantum Hall regime at Landau level filling factor $\nu =1$ is a strong
ferromagnet with total spin quantum number $S = N/2$.
(Here $\nu \equiv N/N_{\phi}$ is the ratio of
the number of electrons to the orbital degeneracy of a Landau level;
$N_{\phi} = A B /\Phi_0 = A / (2 \pi l_{c}^{2})$ where $A$ is the
area of the system, $\Phi_0$ is the magnetic flux quantum, and B is the
magnetic field strength.)  Recent NMR studies\cite{BTPW94} of this
system have provided evidence for the existence of electrically
and topologically charged \cite{BTPW94} Skyrmion excitations
in the ground state for $\nu \ne 1$, as predicted by earlier theoretical
work \cite{LK90}.  These observations have motivated further experimental
studies of Skyrmions based on transport \cite{SEPW95} and optical \cite{AGB95}
measurements.  Here \cite{previous} we address the accurate
NMR measurements \cite{BTPW94} of the temperature dependence
of the spin magnetization, $M(T)$ at $\nu =1$.
These data present an
important challenge to theory since, as we explain at greater
length below, their explanation requires a consistent treatment of
fermion--quasiparticle and collective--magnetization excitations of
the ferromagnetic ground state.  The difficulties posed by this
necessity have, for nearly seventy years, confounded attempts to
develop the early work of Bloch \cite{Blo29} into a definitive theory of
itinerant
electron magnetism\cite{IEF}.  The existence of ferromagnetism in
two--dimensional electron systems provides an important opportunity
for progress, since theory does not need to contend with complicated
band structures which have no essential importance but frustrate
attempts to compare even relatively simple theoretical approximations
with experiment.  In this Letter we report on a theory for
$M(T)$ which employs the simplest credible approximation
in a many--body perturbation theory approach\cite{RS95}.
The absence of band--structure
complications allows many steps in the calculation to be completed analytically
and some relationships between the fermion--quasiparticle and collective
pictures
of the magnetization to be clearly established.  In particular we show
explicitly (for the first time as far as we are aware) that
the magnetization suppression due to thermally excited spin--waves,
which dominates in some limits, appears in a fermion-particle
description as a reduction in the spectral weight of the quasiparticle
pole.

\newpage

The microscopic Hamiltonian for the system in question is:
\begin{eqnarray}
H & = & - \frac{1}{2} \Delta_{z}(N_{\uparrow}-N_{\downarrow})
 +  \frac{1}{2}\; \frac{e^{2}}{(4\pi \epsilon \l_c)} \times \nonumber \\
& \times  & \sum\limits_{p,p^{'},q\not= 0 \atop \sigma, \sigma^{'}}
\tilde{W}(q,p-p^{'}) c^{\dagger}_{p+\frac{q}{2},\sigma}
c^{\dagger}_{p^{'}-\frac{q}{2}, \sigma^{'}}
c^{\vphantom{\dagger}}_{p^{'}+\frac{q}{2}, \sigma^{'}}
c^{\vphantom{\dagger}}_{p-\frac{q}{2}, \sigma} \;.
\label{H}
\end{eqnarray}
Here, the $c^{(\dagger)}_{k,\sigma}$ is a fermion creation operator for the
lowest Landau level in a Landau gauge,
(we choose $\sigma=\uparrow$ as the spin majority direction)
and $N_{\sigma}$ is the number operator for spin $\sigma$.
The two--particle matrix element
for an isotropic interaction projected onto the lowest Landau level is
\begin{equation}
\tilde{W}(q,p-p^{'})= \int \frac{d^{2}\vec {k}}{(2\pi)^{2}}
\tilde{V}(\vec{k}) e^{-\frac{k^{2}}{2}}
e^{i k_{x}(p-p^{'})} \delta_{k_{y},q} \; .
\end{equation}
where $\tilde{V}(\vec{k})$ is the electron-electron interaction.
We emphasize that this Hamiltonian is exact apart from corrections due
to Landau level mixing, which become unimportant at strong magnetic fields.

The two energy scales in $H$ are the Zeeman energy $\Delta_{z} \equiv |g
\mu_{B}B|$
and the electron--electron interaction energy scale
$\lambda=e^{2}/(4\pi \epsilon \l_c)$.
The ratio of these energy scales in the NMR \cite{BTPW94} experiments ($\nu=1$
at $B=7T$), $\Delta_{z}/\lambda \simeq 2.2K/136K = 0.016$, is small.
Because of this fortunate experimental fact, Zeeman coupling to the spins
acts like a weak symmetry breaking field even though
we are in the strong magnetic field limit for the electron's orbital
degrees of freedom.  With the Zeeman field the ground state,
denoted by $|0>$, is the nondegenerate $S_z=N/2$ state
with all spin $\uparrow$ one--particle orbitals occupied\cite{GM95}. As in the
case of a localized--electron ferromagnet the single spin--wave
states ({\it i.e.}, the eigenstates with $S_z = N/2 -1$)
of this system can be determined exactly: \cite{BIE81,KH84,RM86}
$|\vec{k}>= 1/\sqrt{N} \sum_{q} e^{iqk_x} c_{q,\downarrow}^{\dagger}
c_{q-k_y,\uparrow}^{\vphantom{\dagger}}|0>$ and
$\epsilon_{SW}(\vec{k}) = \Delta_{z} + \lambda  (\tilde{a}(0) -
\tilde{a}(\vec{k}))$.
(Here, $\tilde{a}(\vec{k}) = \int (d^2\vec{q}/(2\pi)^2) \tilde{V}(\vec{q})
e^{-q^2/2}e^{i \vec{q} \cdot \vec{k}}$.)
Because of the itinerant character of the electrons
the wavevectors labelling the spin--wave states are
not restricted to a Brillouin--zone;
Pauli blocking of the electronic orbital degrees of freedom becomes
less and less restrictive as more spins flip.  The main importance of
the Zeeman coupling, as detailed below, is to introduce a spin--wave gap
and therefore cutoff their magnetization suppression so that $M$ is finite
at $T \ne 0$.  The challenges in calculating $M(T)$ closely
parallel \cite{fb} those for any itinerant electron ferromagnet \cite{IEF}.

A useful point of reference is the Hartree--Fock approximation (HFA),
\cite{Blo29},
which is
especially simple in the quantum Hall regime \cite{AU74}.
This is the analog for quantum Hall systems of the band theory of
itinerant electron magnetism.
The HF orbital energies measured from the chemical potential solve
the equation $\xi^{HF}_{\downarrow} =- \xi^{HF}_{\uparrow} =  (\Delta_z/2 +
\lambda \tilde{a}(0) (\nu^{HF}_{\uparrow}-1/2))$ where
$\nu_{\uparrow}^{HF}=n_F(\xi^{HF}_{\uparrow})$.  The solutions are
plotted in Fig.~1.
The `exchange enhanced' spin--splitting of the HFA and
the associated enhancement of the spin--polarization over its single
particle value persist to high--temperatures.  The sharp inflection points in
Fig.~1 are a remnant of the spontaneous ($\Delta_z=0$)
magnetization which occurs {\it incorrectly} in the HFA
for $k_B T <  k_B T^{HF}_c \equiv \tilde{a}(0)/4$.  The well known failure of
HF theory for itinerant electron ferromagnets \cite{IEF} is particularly
stark in the present case.

The weakness of the HFA rests in its
inability to account for collective excitations of the ferromagnet
which play a dominant role at low temperatures if $\Delta_z$ is small.
In a many--body perturbation theory approach the simplest approximation which
reflects the presence of spin--wave excitations in itinerant ferromagnets is
one which includes a self--energy insertion consisting of a
ladder sum of repeated interactions between HF electrons of
one spin and holes of the opposite spin.  The corresponding approximation has
been discussed previously in theories of Hubbard model systems \cite{edwards}.
In the case of
quantum Hall ferromagnets the sum may be evaluated explicitly and we
find for the majority spin \cite{later}:
\begin{eqnarray}
& & \Sigma_{\uparrow}(i \omega_{n} )   =  \lambda^{2}(\nu_{\uparrow}^{HF}
- \nu_{\downarrow}^{HF}) \times  \nonumber \\
&  \times & \int_{0}^{\infty}
d(\frac{k^2}{2}) \tilde{a}^{2}(k)
\frac{\lbrace n_{B}(\tilde{\epsilon}_{SW}(k)) + \nu^{HF}_{\downarrow}
)\rbrace }{(i\hbar \omega_{n} + \tilde{\epsilon}_{SW}(k) -
\xi_{\downarrow}^{HF})}.
\label{SE1}
\end{eqnarray}
A similar result is obtained for the minority spin.  In Eq.~(\ref{SE1})
$n_{B}(\tilde{\epsilon}(k))$ is the Bose-Einstein distribution function
and
\begin{equation}
\tilde{\epsilon}_{SW}(\vec{k})  =  \Delta_z + \lambda (\nu_{\uparrow}^{HF} -
\nu_{\downarrow}^{HF})
(\tilde{a}(0) - \tilde{a}(\vec{k}))
\label{SW1}
\end{equation}
is the finite--temperature spin--wave dispersion in this approximation.
This self--energy describes the emission and absorption of a virtual
spin--wave;
$\tilde{a}(k)$ which describes the spin--wave dispersion is,
significantly as we discuss below,
also the coupling constant for the interaction of electrons with spin--waves.

After analytical continuation ($i \hbar \omega_n \to \hbar \omega + i \eta $)
we see that the $\Sigma^{ret}_{\uparrow}$ is complex in the
interval $I \equiv (\xi^{HF}_{\uparrow},\xi^{HF}_{\downarrow}-\Delta_z)$,
where real transitions with electronic spin flips and
spin--wave absorption are possible, and is real outside $I$.  The real
part of $\Sigma^{ret}_{\uparrow}$ diverges to $-\infty$ at the lower limit of
$I$ and to $+\infty$ at its upper limit so that the retarded
Green's function $G^{ret}_{\uparrow}(\omega) = (\hbar \omega -
\xi^{HF}_{\uparrow} -
\Sigma^{ret}_{\uparrow}(\omega))^{-1}$ has a branch cut along $I$ and poles on
opposite sides at $\omega^{*}_{\pm}$.
Note that $\Sigma^{ret}_{\uparrow}$ vanishes and the HF results
are correctly recovered for $T \to 0$.  For $T \to \infty$ $\Sigma^{ret}
_{\uparrow}$
again vanishes because $\nu^{HF}_{\uparrow}-\nu^{HF}_{\downarrow}$
vanishes.  In the intermediate temperature region corrections
to the HFA are important.

To compare with the NMR $M(T)$ measurements we evaluate the spin
magnetization from $A_{\uparrow}(\omega)=  - 2 Im
G^{ret}_{\uparrow}(\omega)$, using
$M(T) = (M_{0}/\nu) \int (d \omega/ 2 \pi) n_{F}(\omega)
(A_{\uparrow}(\omega)-A_{\downarrow}(\omega))$
and $A_{\uparrow}(\omega)= A_{\downarrow}(-\omega)$.  (The
last identity is a consequence of particle--hole symmetry.)
Results \cite{screening} for $A_{\uparrow}(\omega)$ are shown in Fig.~2 where
we see that deviations from the single $\delta$ function peak of
the HFA increase strongly with temperature.
At low $T$ nearly all of the spectral weight is in the
pole at $\omega^{*}_{-}$ which is shifted only slightly from
$\xi^{HF}_{\uparrow}$.
For $T\to 0$ we see from Eq.~(\ref{SE1}) that the residue of this pole is
given by
\begin{eqnarray}
z_{-} &\to& 1 + \frac{\partial \Sigma
_{ \uparrow}^{ret}(\omega)}{\partial \omega}_{|\omega = \xi^{HF}_{\uparrow}|}
\nonumber \\
& \to & 1 - \int_{0}^{\infty}
d(\frac{k^2}{2}) n_{B}(\tilde{\epsilon}_{SW}(k)).
\label{residue}
\end{eqnarray}
In the last form of Eq.~(\ref{residue}) we have set $\nu^{HF}_{\downarrow}=0$
for $T \to 0$.  Note that $\tilde{a}(k)$ cancels
out of the final form of Eq.(~\ref{residue}) because of the
relationship between the coupling function and the spin--wave dispersion.
 From this result we see that the reduction in $M$ at low $T$ due
to thermally excited spin--waves, familiar from continuum
and Heisenberg type models of magnetism (see for example Ref.
\cite{RS95}) appears in a microscopic itinerant electron
theory as a reduction in the quasiparticle normalization factor
on going beyond a HFA.

Results for $M(T)$ calculated from these spectral functions
for a set of parameters appropriate to the NMR experiments\cite{BTPW94}
are shown in Fig.~3 and compared with results from other theories.
The HFA yields $M(T)/M_{0} = 2n_{F}(\xi_{\uparrow}^{HF}) - 1 =
\tanh(\beta \xi_{\downarrow}^{HF}(\beta)/2))$ and is exact both
for $T \to 0$ and $T \to \infty$.  As expected it grossly overestimates
the observed magnetization.  The present theory
predicts a strong suppression of the magnetization at much lower temperatures
because it includes the effects of thermally excited spin--waves.
Our results are extremely insensitive to the choice of screening
wavevector\cite{screening}.
A more complete theory would have temperature and frequency dependent
screening; the value $q_{sc}=0.01 l_c^{-1} $,
is the\cite{later} self--consistent static screening
wavevector temperature in our theory for $T\simeq 0.09$.
We observe that an excellent approximation to our results
is obtained by including only the two poles in the spectral function.
The two poles correspond respectively to the Hartree--Fock quasiparticle state
and a state with a minority spin quasiparticle and a spin--wave.  This
suggests that at higher temperatures an improved theory would need
to describe multiple spin--wave dressings of the fermion propagator.
At low temperatures the reduction in $M(T)$ is dominated by
the long--wavelength spin--wave contribution which gives
$ M(T)/M_{0}-1  = C(T) T ln (1 - e^{-\Delta_z/k_B T})$ where $C(T)$ depends
weakly on temperature.  This suppression crosses over from an
activated temperature dependence
( $- C(T)T e^{-\Delta_z/k_{B}T}$) for $k_B T < \Delta_z$ to an
approximately linear temperature
dependence ($-C(T)T ln(\Delta_z/k_BT)$)
for $k_B T > \Delta_{z}$ but still small.
The linear $T$--dependence is the two--dimensional analog of the
$T^{3/2}$--Bloch law familiar in three--dimensional ferromagnets \cite{Mat88}.

To compare with experiment it is necessary to account for the finite
thickness of the two--dimensional electron layer by including form
factors in determining the effective electron--electron interaction.
We see in Fig.~3 that although our theory overestimates
$M(T)$, the improvement compared to the HFA is considerable.
One obvious shortcoming of the simple theory is the use of HFA
propagators in calculating $\tilde{\epsilon}_{SW}(k)$; the fact that this
quantity depends on the propagator only through $\nu^{HF}_{\sigma}$
suggests that the theory could be improved simply by replacing
the HF filling factors by the those calculated including
self--energy corrections.  As shown in Fig.~3, applying this procedure
leads to an further decrease of $M(T)$.  Overall the agreement of
this relatively simple theory with experiment gives hope that a
fairly complete theory can be built from this starting point.
It is possible that part of the discrepancy originates in the
difficulty of precisely locating $\nu =1$ experimentally,\cite{BTPW94},
since small changes in filling factor are known to
cause dramatic changes in the ground state magnetization.
However, we believe that most of the discrepancy in Fig.~3 is
due to limitations of the theory.  The opportunity
for such a direct comparison of theory and experiment should
be helpful in developing a more complete theory.  We
believe that any lessons learned from advances in this direction
will be transferable to all itinerant electron ferromagnets.

We thank S.~Barrett for discussions and for sharing
data from the NMR--measurements
prior to publication.  Helpful interactions with W.~Apel, S.~M.~Girvin,
C.~Hanna, R.~Haussmann and H.~Mori are gratefully acknowledged.
One of the authors (M.K.) is supported by a fellowship from the
German Academic Exchange
Service (DAAD). This work was supported in part by NSF grant No.~DMR94-16906.
\\

\noindent
$^{\dagger}$ Present address: Institut f\"ur Theoretische Physik,
Universit\"at des Saarlandes, \newline
PF 15 11 50, D--66041 Saarbr\"ucken, Germany.

\newpage
\noindent
{\bf Figure captions:} \\

{\bf Fig.~1:}
HFA orbital energies $\xi_{\sigma}^{HF}$ in units of
$\lambda$ as a function of temperature for Coulombic electron--electron
interactions and $\nu =1$.  The solid line is $\xi^{HF}_{\uparrow}$
and the long--dashed line is $\xi^{HF}_{\downarrow}$.  These results are for
$\Delta_z=0.016$.  The HFA spin magnetization is
$M = M_{0}(\nu^{HF}_{\uparrow} - \nu^{HF}_{\downarrow})$ where
$M_{0}= (|g\mu_{B}|/2)N$ is the total magnetization at $T=0$. \\

{\bf Fig.~2:}
The spectral function $A_{\uparrow}(\omega)$ for parameter values
$\Delta_z = 0.016$ and $q_{sc} l_c = 0.01$ at temperatures
$T=0.05, 0.1$ and $0.2$ in units of $\lambda / k_B$ ($\Delta_z$
and $\omega$ are in units of $\lambda$).
The numbers indicate the fraction of the total spectral weight from the
two poles and from the branch cut. \\

{\bf Fig.~3:}
Results for $M(T)$ for $\Delta_z=0.016$.
The long--dashed and short--dashed lines show the results of the
present theory with two-different choices for the {\it ad hoc}
screening vector, $q_{sc}=0.01 l_c^{-1}$ and $q_{sc} = 0.1 l_c^{-1}$.
Results incorporating a finite quantum well width ($w = 30 nm \simeq 3.11 l_c$)
with $q_{sc}=0.01 l_c^{-1}$ are shown as a solid line.
The experimental results of
Barrett {\it et al.} are shown as crosses.
For comparison, the dot--dashed curve is the result
of the SC--HFA ($q_{sc}=0.01 l_c^{-1}$) theory and the solid curve with stars
shows the magnetization obtained when the spin-wave dispersion is
calculated from self-consistently determined partial filling factors.


\begin{thebibliography}{99}
\bibitem{later} A detailed description of this calculation will
be published later.
\end{thebibliography}
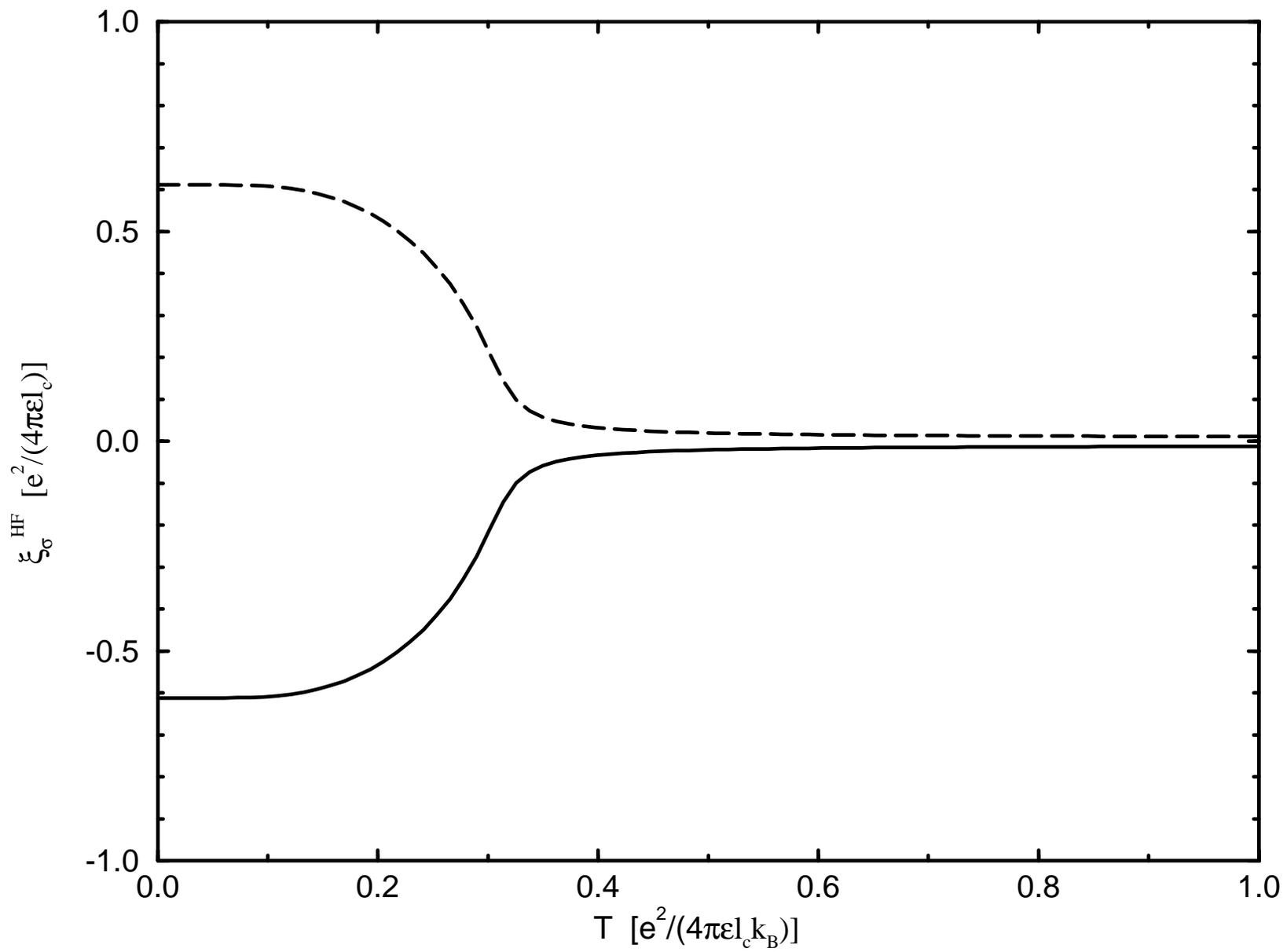
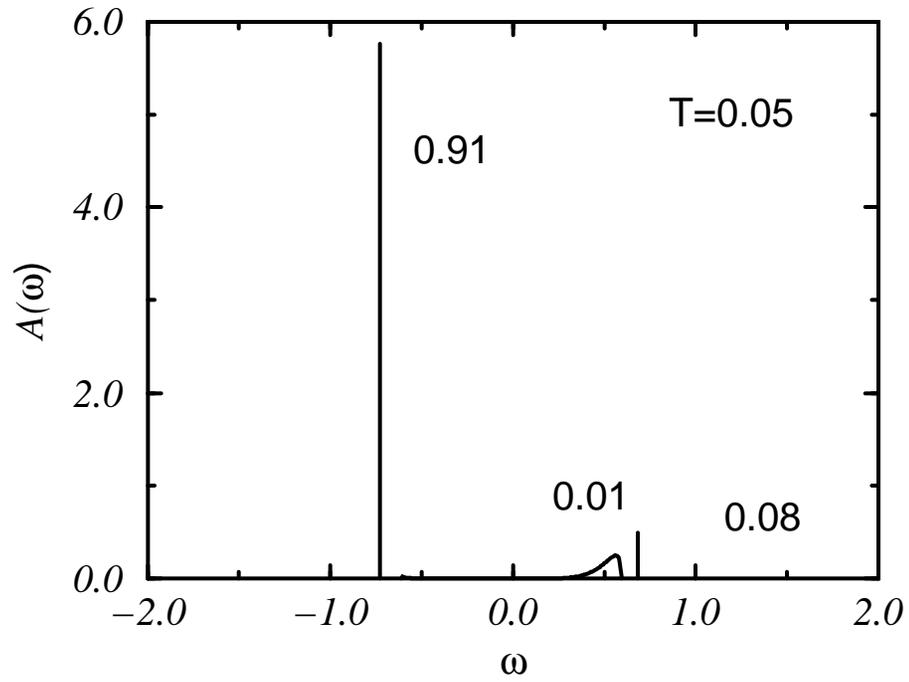
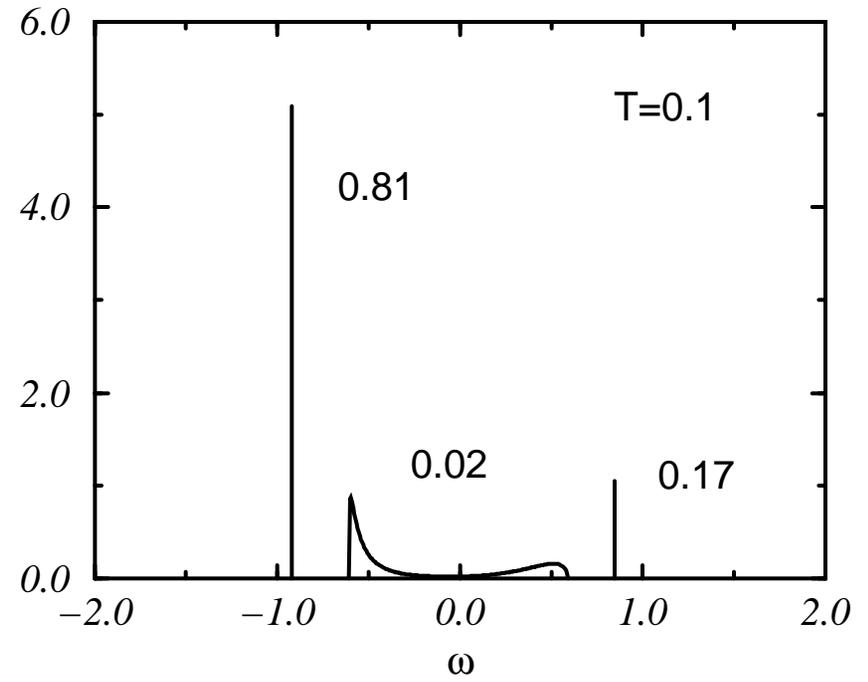
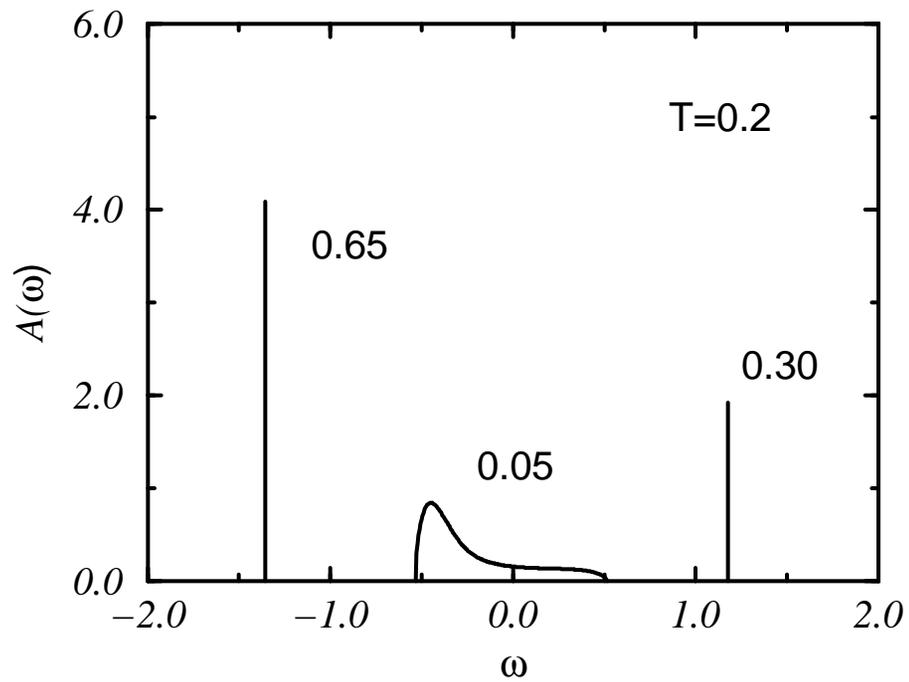
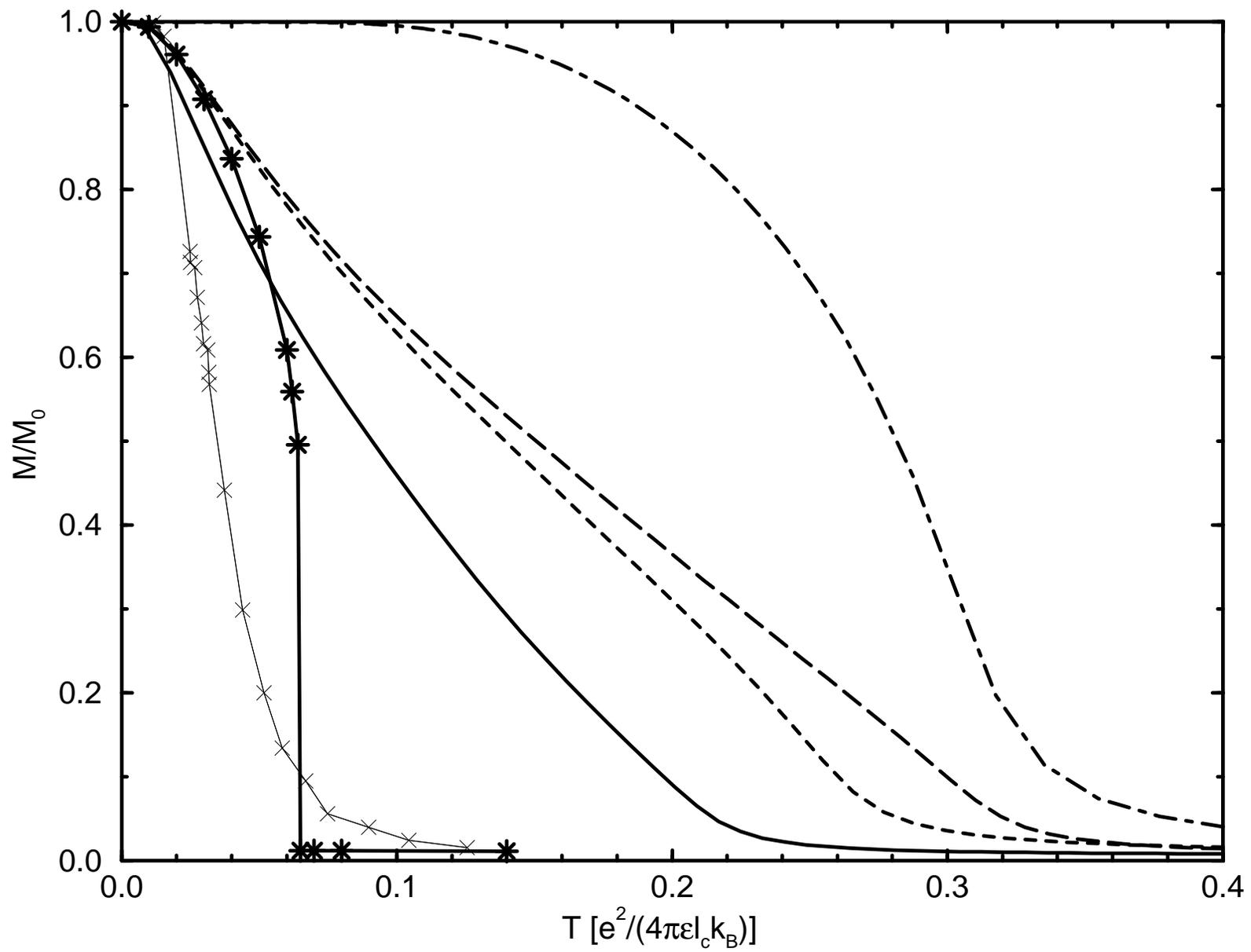
\end{document}